\documentclass[fleqn,twoside]{article}
\usepackage{espcrc2}

\usepackage{graphicx}
\usepackage[figuresright]{rotating}

\newcommand{\AmS}{{\protect\the\textfont2
  A\kern-.1667em\lower.5ex\hbox{M}\kern-.125emS}}

\title{Soft data and the hard pomeron}

\author{J.R. Cudell\address{JR.Cudell@ulg.ac.be;
Institut de Physique, 
Universit\'e de Li\`ege, 
4000 Li\`ege, Belgium },
A. Lengyel\address{sasha@len.uzhgorod.ua;
Institute of Electron Physics, Universitetska 21, UA-88000
Uzhgorod, Ukraine.},
E. Martynov\address{martynov@bitp.kiev.ua;
Bogolyubov Institute
for Theoretical Physics, 03143 Kiev, Ukraine}, and
O.V. Selyugin\address{selugin@qcd.theo.phys.ulg.ac.be; Institut de Physique,
Universit\'e de Li\`ege,
4000 Li\`ege, Belgium}\thanks{on leave from the Bogoliubov Theoretical
Laboratory, JINR, 141980 Dubna, Moscow Region, Russia.}
}       
\begin{document}

\begin{abstract}
We show that the introduction of an additional, hard, singularity 
in soft forward amplitudes enables one to improve considerably 
the fits using a simple-pole structure for the soft pomeron.
\vspace{1pc}
\end{abstract}

\maketitle

\section{Introduction}

One of the central questions in QCD concerns the continuation of
short-distance results to the soft region or, in other words, the
extension of the large-$Q^2$ perturbative description of the data 
($F_2$, $F_2^c$,...) to the non-perturbative $Q^2=0$ domain 
($\sigma_{tot}$, $\rho$). 
There are several phenomenological models that already
attempt to describe such a transition \cite{DoDo,BFKLunit}. 
Most of them assume a rise in total cross
sections proportional to $\log^2(s/s_0)$, which describes best
the data, both for $\sigma_{tot}$ and $\rho$~\cite{COMPETE}. Such
a behaviour in $s$ $-$ or $1/x$ $-$ can be used as an initial condition
for evolution equations~\cite{Soyez1}, and is even a very good
approximation to the result of the evolution~\cite{Soyez2}. Such
a rise can be obtained via a strong unitarisation of a BFKL-like
singularity~\cite{BFKLunit}.

Following~\cite{paper}, we want here to explore another possibility: we shall assume that ``bare'' exchanges of Regge trajectories account for most of the elastic amplitude, and that unitarisation and cuts have a small, negligible, 
effect. This is the standard assumption that lead Donnachie and Landshoff~\cite{DoLa} to a very simple and successful description of soft
scattering, based on the exchange of two degenerate leading 
meson trajectories and of a soft pomeron. Recently, it was
however found that such a simple singularity
structure does not allow a good description neither of soft forward data~\cite{COMPETE} nor of $F_2$. In the latter case, Donnachie and 
Landshoff~\cite{DoLahard} have shown that the introduction of a second pomeron with a large intercept $\alpha_h\approx 1.4$ is consistent 
both with the evolution equations and with the data. We want here
to address the former question, and show that the introduction
of the $same$ hard singularity makes the fit to $\sigma_{tot}$
and $\rho$ as good as that based one a $\log^2 (s/s_0)$ rise.

\section{Improved analysis}
\begin{figure}
\resizebox*{0.4\textwidth}{!}{\includegraphics{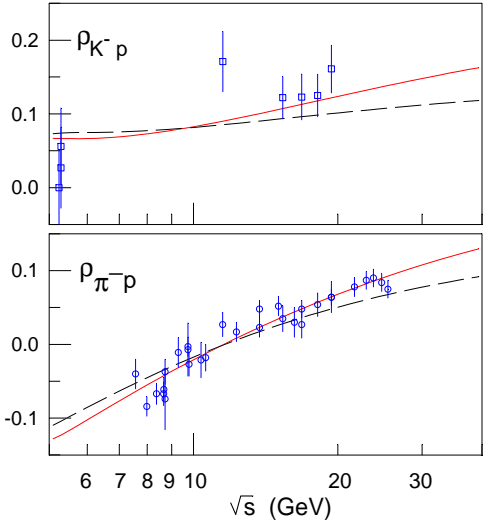}}
%\\ \resizebox*{0.3\textwidth}{!}{\includegraphics{rhopiKp.ps}} 
\caption{Difference between \( \rho  \) values fitted with (plain)
and without (dashed) a hard pomeron, assuming all singularities are
simple poles.}
\end{figure}
In the analysis of ref.~\cite{COMPETE}, models based on a simple-pole
pomeron were dismissed mainly because they could not describe the
real part of the amplitude well, or equivalently because the fit
to 
$\rho$, the ratio of the real part to the imaginary part of the elastic hadronic amplitude, had an
unacceptably high $\chi^2$. Hence we have improved the treatment
of the real part of the amplitude by introducing the following 
refinements:\\
$\bullet$ We include and fit the subtraction constant present in the
real part of the amplitude because of rising $C=+1$ contributions;
\\ $\bullet$ We use integral dispersion relations down to the correct threshold and, at low energies (for which the analytic asymptotic model
is not correct), we use (a smooth fit to) the data for $\sigma_{tot}$ 
to perform the dispersion integral. \\
$\bullet$ We use the exact form of the flux factor ${\cal F}=2m_p p_{lab}$
and Regge variables $\tilde s\equiv {s-u\over 2}$ proportional to
$\cos(\theta_t)$ instead of their dominant terms at large-$s$\footnote{In the $\gamma\gamma$ case, we use ${\cal F}=s$}. 

Following~\cite{COMPETE}, we fit total cross sections and $\rho$
for $pp$, $\bar p p$, $\pi^\pm p$ and $K^\pm p$, and total cross sections
for $\gamma p$ and $\gamma\gamma$ in the region 5 GeV $\leq\sqrt{s}$.
Furthermore, as we are using simple poles, we use 
Gribov-Pomeranchuk factorisation of the
residues  
at each simple pole to predict the $\gamma\gamma$ amplitude
from the $pp$ and $\gamma p$ data~\cite{CMS}. 

If we define the hadronic $ab$ amplitude as ${\cal A}_{ab}={\Re}_{ab}+i{\Im}_{ab}$, we obtain the total cross section as
\begin{equation}
\sigma _{tot}^{ab}\equiv \frac{1}{2 m_b p_{lab}}\, \Im^{ab}
\end{equation}
with $p_{lab}$ the momentum of particle $b$ in the $a$ rest frame.
The real part of the amplitude can then be obtained from integral dispersion relations:
\begin{eqnarray}
\label{eq:final dr}
\Re_{ab}&=& R_{ab}\nonumber\\
&+&\frac{E}{\pi}\ {\textrm{P}}\int
_{m_{a}}^{\infty }\left[ \frac{\Im_{ab }}{E'-E}-\frac{\Im_{a \bar b
}}{E'+E}\right]\, {dE'\over E'},
\end{eqnarray}
where \( E \)
is the energy of $b$ in the rest frame of $a$, P indicates that we have to
do a principal-part integral, and \( R_{ab} \) is the subtraction constant.

The models that we consider are defined by the following equation:
\begin{equation}
\label{fluxfac}
{\Im}_{ab}\equiv s_1\left[ {\Im}^{R+}_{ab}\left({\tilde s\over s_1}
\right) +{\Im}^{S}_{ab}\left( {\tilde s\over s_1}\right) \mp {\Im}^-_{ab}
\left({\tilde s\over s_1}\right) \right] ,
\end{equation}
with $s_1=1$ GeV$^2$, and the $-$ sign in the last term for particles.
For the two reggeon contributions, we use simple-pole expressions 
${\Im}^{R+}_{pb}=P_{b}\left({\tilde s\over s_1}\right)^{\alpha _{+}}$ 
and ${\Im}^-_{pb}=M_{b}\left({\tilde s\over s_1}\right)^{\alpha _-}$. For the pomeron
contribution, we allow two simple poles to contribute:
\begin{equation}
\label{poles}
\Im^S_{pb}=S_{b}\left( {\tilde s\over s_1}\right)^{\alpha _{o}}
+H_{b}\left({\tilde s\over s_1}\right)^{\alpha _{h}}
\end{equation}
For comparison, we also consider expressions corresponding to
a dipole or a tripole:
\begin{equation}
\label{dipole}
\Im^S_{pb}=D_{b}{\tilde s\over s_1}\ln \frac{\tilde s}{s_{d}},
\end{equation}
\begin{equation}
\label{tripole}
\Im^S_{pb}=T_{b}{\tilde s\over s_1}\left[\ln ^{2}\frac{\tilde s}{s_{t}}+t'_{b}\right].
\end{equation}
\begin{figure}
\resizebox*{0.4\textwidth}{!}{\includegraphics{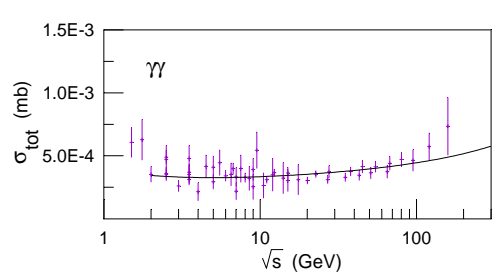}} 
\caption{Fit to the $\gamma\gamma$ total cross section using Gribov-Pomeranchuk factorisation.
}
\end{figure}

\section{Results}
The improved treatment of $\rho$ leads to a better fit in all cases (a
dipole pomeron (\ref{dipole}) reaches a $\chi^2/dof$ of 0.94 and
a tripole pomeron (\ref{tripole}) 
one of 0.93, whereas they were both 0.98 in the 
standard analysis \cite{COMPETE}).
However, if we use only one simple pole for the pomeron 
($i.e.$ if we set $H_b=0$ in
(\ref{poles}), we still cannot get a fit comparable to those obtained using (\ref{dipole}, \ref{tripole}).
~\\

{\centering \begin{tabular}{|c|c|}
\hline
Parameters & value  \\ \hline\hline
\( \alpha _{h} \) & 1.45 $\pm$  0.01 \\ \hline
\( H_{p} \)& 0.10 $\pm$  0.02 \\ \hline
 \( H_{\pi } \)& 0.28 $\pm$  0.03 \\ \hline
\( H_{K} \)& 0.30 $\pm$  0.03 \\ \hline
\( H_{\gamma } \)& 0.0006 $\pm$  0.0002 \\ \hline\hline
\( \alpha _{o} \) & 1.0728 $\pm$ 0.0008 \\ \hline
\( S_{p} \)& 56.2 $\pm$ 0.3 \\ \hline
\( S_{\pi } \) & 32.7 $\pm$ 0.2  \\ \hline
\( S_{K} \)& 28.3 $\pm$  0.2  \\ \hline
\( S_{\gamma } \)& 0.174 $\pm$  0.002  \\ \hline\hline
\( \alpha _{+} \) & 0.608 $\pm$  0.003 \\ \hline
\( P_{p} \)& 158 $\pm$  2 \\ \hline
\( P_{\pi } \)& 78 $\pm$  1 \\
\hline \( P_{K} \) & 46 $\pm$  1 \\ \hline
\( P_{\gamma } \)& 0.28 $\pm$  0.01 \\ \hline\hline
\( \alpha _{-} \) & 0.473 $\pm$  0.008 \\ \hline
\( M_{p} \)& 79 $\pm$  3 \\ \hline
\( M_{\pi } \)& 14.2 $\pm$  0.5 \\ \hline
\( M_{K} \)& 32 $\pm$  1 \\ \hline\hline
\( R_{pp} \)& -164 $\pm$  33 \\ \hline
\( R_{p\pi } \)& -96 $\pm$  21 \\ \hline
\( R_{pK} \)& 3 $\pm$  26 \\ \hline \end{tabular}\par}

{\noindent Table 1\\ Parameters obtained in the fit from 5 to 100 GeV.}
~\\

However, we found that the inclusion of the second singularity in 
(\ref{poles}) has a dramatic effect: the $\chi^2$ drops from 661 to 551 for
619 points, nominally a 10 $\sigma$ effect! More surprisingly, the
new singularity has an intercept of 1.39, very close to that 
obtained in DIS by Donnachie and Landshoff. However, as was already known 
\cite{pre-compete}, the new trajectory, which we shall call the hard pomeron,
almost decouples from $pp$ and $\bar p p$ scattering. Nevertheless, it improves
considerably the description of $\pi p$ and $K p$ amplitudes, and 
parametrisation (\ref{poles}) becomes as good as (\ref{tripole}).

The decoupling in $pp$ and $\bar p p$ scattering can easily be understood: any sizable
coupling will produce a dramatic rise with $s$, and only $pp$ and $\bar p p$ 
data reach high energy. For these data, the hard pomeron contribution will surely need to be unitarised (see however~\cite{DoLah} for
a different opinion). To get a handle on the hard pomeron parameters, it is thus a good idea to fit to lower energies first. We choose to consider the region 
from 5 to 100~GeV (which includes all the $\pi p$ and $Kp$ data). We checked
that the parameters describing the hard pomeron component are stable if we 
slightly change the region of interest, by augmenting the minimum 
energy to $e.g.$ 10 GeV, or by decreasing the maximum energy to 
$e.g.$ 40 GeV.

%\section{Result for 100 GeV $\geq \sqrt{s}\geq 5$ GeV}
Hence our best estimate for the parameters describing the hard pomeron is shown in Table 1, and some of the fits are shown in Fig.~1.

As can be seen, the parameters are in agreement with those describing DIS, 
although the soft pomeron is a bit softer than assumed in~\cite{DoLahard}.
In particular, we obtain for the hard-pomeron singularity
\begin{equation}
\alpha_h=1.45\pm 0.01.
\end{equation}
However, a new and unexpected hierarchy of couplings is needed. Writing
$H_{ab}=h_a\ h_b$, we obtain:
\begin{equation}
1\approx h_K\approx 1.1 h_\pi\approx 3.2 h_p\label{coups} 
\end{equation}
Such a hierarchy may be expected in dipole models~\cite{DoDo}, but it is
stronger than expected. Size effects may thus not be sufficient 
to account for it.

The hard pomeron is probably not a simple pole, but it must be close to it: as we obtain the $\gamma\gamma$ cross section via Gribov-Pomeranchuk factorisation, we indeed test the analytic nature of the singularity \cite{CMS}. 
The result is shown in Fig.~2.
One can see that the LEP data are compatible with our results, 
and that we prefer a lower value, such as that obtained using PHOJET.
Also, our value of the coupling of the hard pomeron to protons must
be an upper limit: bigger values would lead to too small 
a $\gamma\gamma$ cross section.
\begin{figure}
{\resizebox*{0.4\textwidth}{!}{\includegraphics{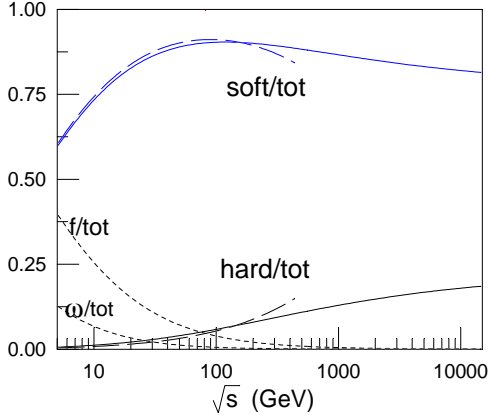}} }
\caption{Relative contribution of the various terms of the amplitude,
compared with the \( C=+1 \) part of the amplitude ({}``tot{}'')
in the \( pp \) case. The dashed curve is for a hard pole, and the
plain curves for the unitarised form.}
\end{figure}

%\section{Extension to higher energies}
To obtain a fit to higher energies, one must surely unitarise the hard
pomeron contribution, as it violates the black-disk limit around $\sqrt{s}=400$ GeV. The way to do this is far from clear, especially as there can be some mixing with the other trajectories.
We have shown in~\cite{paper} that it is possible to find a unitarisation scheme
which produces a good description of the data for all energies. 
Its contribution, shown in Fig.~3, is always smaller than 25\% of the total
cross section. 

\section{Outlook}
In conclusion, we have shown \cite{paper} that it is possible that the hard pomeron is present in soft data
and that this object may be similar to a simple pole for $\sqrt{s}\leq 100$ GeV.
This means that the bulk of the simple-pole phenomenology
can be kept in the presence of a hard pomeron. The latter however will bring
corrections at large $s$, large $Q^2$ or large $t$. 
The surprising hierarchy of couplings (\ref{coups}), as well as the
smallness of the residues, indicates that our results will need confirmation. It may be worth noting here that we obtain similar results
if we exclude the $\rho$ data from the analysis. Also, some
evidence for a hard pomeron may also be found in elastic scattering data \cite{coming}.

\end{document}